\documentclass[runningheads]{llncs}

\usepackage{amssymb}
\usepackage{booktabs} 
\usepackage{multicol} 
\usepackage{tabu}                      

\usepackage{graphicx}

\begin{document}
\title{All Factors Should Matter! Reference Checklist for Describing Research Conditions in Pursuit of Comparable IVR Experiments\thanks{Supported by KOBO Association and XR Lab at PJAIT.}}
\author{Kinga H. Skorupska\inst{1,2}\orcidID{0000-0002-9005-0348} \and
Daniel Cnotkowski\inst{3} \and
Julia~Paluch\inst{1}\orcidID{0000-0002-7657-7856} \and
Rafa{\l} Mas{\l}yk\inst{1}\orcidID{0000-0003-1180-2159}\and
Anna~Jaskulska\inst{4}\orcidID{0000-0002-2539-3934} \and
Monika Kornacka \inst{2}\orcidID{0000-0003-2737-9236} \and
Wies{\l}aw Kope{\'c} \inst{1,2}\orcidID{0000-0001-9132-4171}} 
\authorrunning{Skorupska et al.}
\titlerunning{All Factors Should Matter!}
%
\institute{Polish-Japanese Academy of Information Technology\\
\email{skorupska@pja.edu.pl}\\
\and
SWPS University of Social Sciences and Humanities \and
National Information Processing Institute \and
KOBO Association}
\maketitle              
\begin{abstract}
A significant problem with immersive virtual reality (IVR) experiments is the ability to compare research conditions. VR kits and IVR environments are complex and diverse but researchers from different fields, e.g. ICT, psychology, or marketing, often neglect to describe them with a level of detail sufficient to situate their research on the IVR landscape. Careful reporting of these conditions may increase the applicability of research results and their impact on the shared body of knowledge on HCI and IVR. Based on literature review, our experience, practice and a synthesis of key IVR factors, in this article we present a reference checklist for describing research conditions of IVR experiments. Including these in publications will contribute to the comparability of IVR research and help other researchers decide to what extent reported results are relevant to their own research goals. The compiled checklist is a ready-to-use reference tool and takes into account key hardware, software and human factors as well as diverse factors connected to visual, audio, tactile, and other aspects of interaction.

\end{abstract}
\section{Introduction}

\begin{figure}[tb]
 \centering 
 \includegraphics[width=\columnwidth]{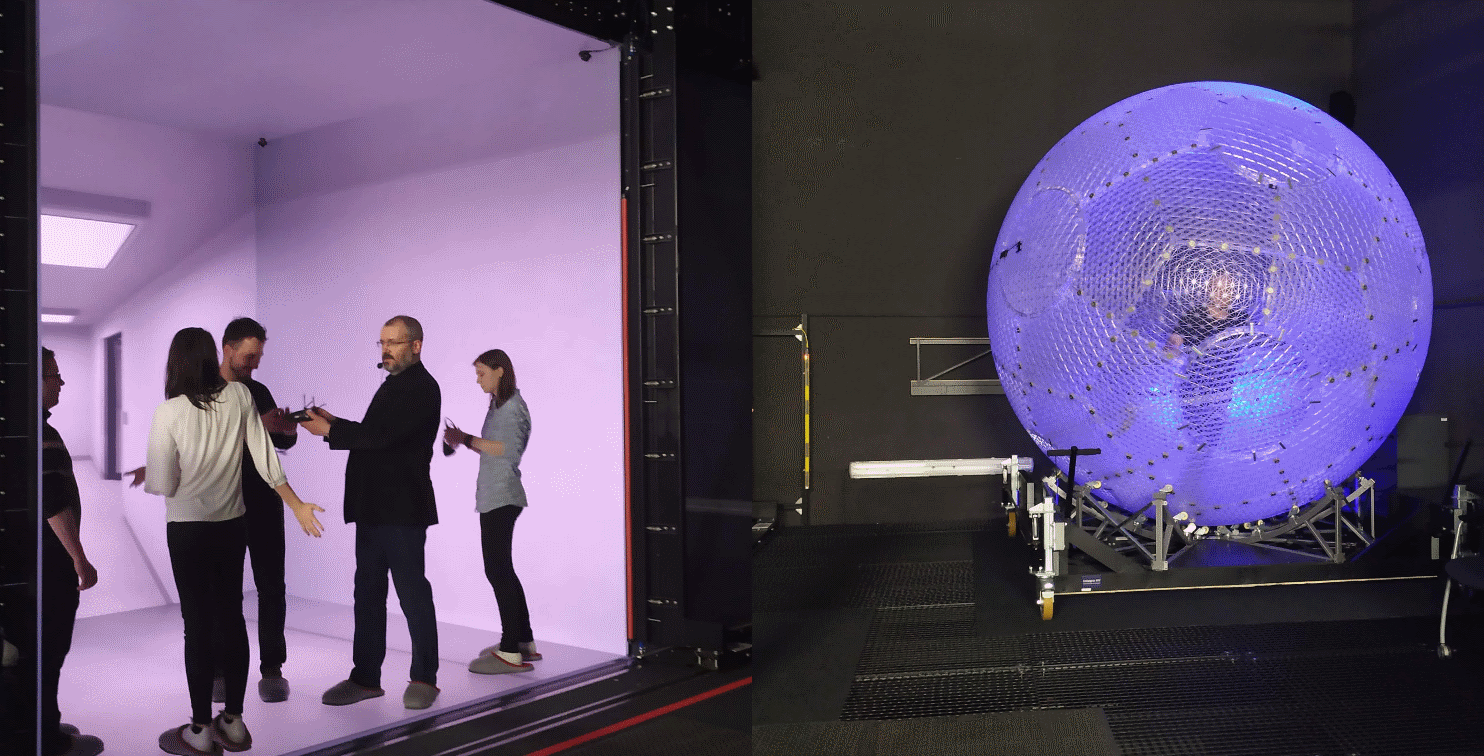}
 \caption{An Immersive Virtual Reality CAVE, where the IVR experience can be shared with others and the "hamster wheel" can simulate covering large distances in VR.}
 \label{fig:cavehamster}
\end{figure}

VR has a long history of development on the reality-virtuality continuuum \cite{RVcontinuum} which may include any system that makes use of artificial elements to create a virtual experience, from the Sensorama machine of 1956, through 3D graphics on a single flat screen or immersive CAVE environments (Figure \ref{fig:cavehamster}) to modern Head Mounted Displays (HMDs).

Consumer solutions were deemed mature enough to conduct scientific experiments as of 2016 \cite{Anthes2016} and have made it possible to perform immersive virtual reality (IVR) studies on a budget. This created an illusion that the experiments conducted with IVR are easily comparable. For example, Buschet al. conclude “virtual environments can be an alternative to real environments for user experience studies, when a high presence is achieved.” \cite{Busch2014} and as such, they can be used to simulate real experiences. 
The excitement of human subjects experiments in IVR environments in some cases overshadowed the human, environmental and technological factors which go hand in hand for creating specific research conditions. Researchers
often omit or ignore some of these factors or at least neglect to mention them in their publications, when it comes to IVR experiments. Such omissions have a lasting effect on the quality of reporting, and the conclusions others can draw based on them.
Andujar and Brunet \cite{Andujar2015} list issues with VR experiment validation related to "lack of background on experimentation, psychology and psychophysics, time-consuming nature of human-subject experiments, and the difficulties to fulfill all requirements (double-blind experiments, informed consent, representative users, representative datasets/models/tasks, reproducibility)". These issues are in part related to the knowledge of IVR experimenters who have to exhibit interdisciplinary competence and take into account both human and technological factors. Andujar and Brunet also noticed that the performance of participants of VR experiments may be related to hardware factors. Continuous development in the field of dimensional tracking, high fidelity displays, wireless transmission of data and motion controls, has brought many improvements over each new generation of previous VR hardware. Some aspects may even be hard to compare across different HMD versions, not only manufacturers, as the specificity of each of them is reflected in the quality of experience they offer. This fast development pace of IVR solutions may cause older research to become obsolete - old Oculus or first Vive HMDs offered an inferior experience to their newest counterparts, but are leaps ahead of Google Cardboard. 
Thus, the validity of conclusions is connected to the specific setup, participants' psychological features (eg. executive function and attention) and domain. This all contributed to the significant problem with IVR experiments which is related to the ability to compare research conditions and, in consequence, the applicability of research results and their contribution to the shared body of knowledge on human factors and IVR. This is a very important limitation which is difficult to overcome, but it can be mitigated.

There are multiple guidelines on good research practices in the European Code of Conduct for Research Integrity\footnote{The code was developed by ALLEA - the European Federation of Academies of Sciences and Humanities and more on this can be found here: https://allea.org/code-of-conduct/}, but such documents are written in broad strokes and do not cover the specifics of each subfield, especially one as complex as the IVR landscape now.  Within it, there is diverse research being conducted from case studies of creating interactive VR archaeological exhibits \cite{Barbieri2018}, to making low-cost VR videos to provide higher level of comfort to pre-surgery patients \cite{OSullivan2018} or coping with stress and regulating emotions \cite{kornacka2016steer}.
Researchers explore various aspects of VR, such as the feeling of presence, connected to minute details like realistic light, sounds or movements \cite{Lee2016}; the sense of agency \cite{Arge2016}, or even attitudes towards entering and exiting VR environments \cite{Knibbe2018}.  

This problem is also directly related to the "reproducibility crisis", one of the driving forces of the open science movement. In psychology research, in response to some of these problems, checklists, such as the one created by Liao et al.\cite{liao_systematic_2016}, started to appear. Software engineering also saw its share of checklists, as in the review by Wieringa \cite{wieringa_lessons_2012}. The same is needed in the area of IVR experiments, which has its own specificity.

Therefore, we have decided to draft an IVR checklist that may guide researchers to report their research in a more comprehensive way. 

\section{What Factors Matter and Why}

\begin{figure}[tb]
 \centering 
 \includegraphics[width=\columnwidth]{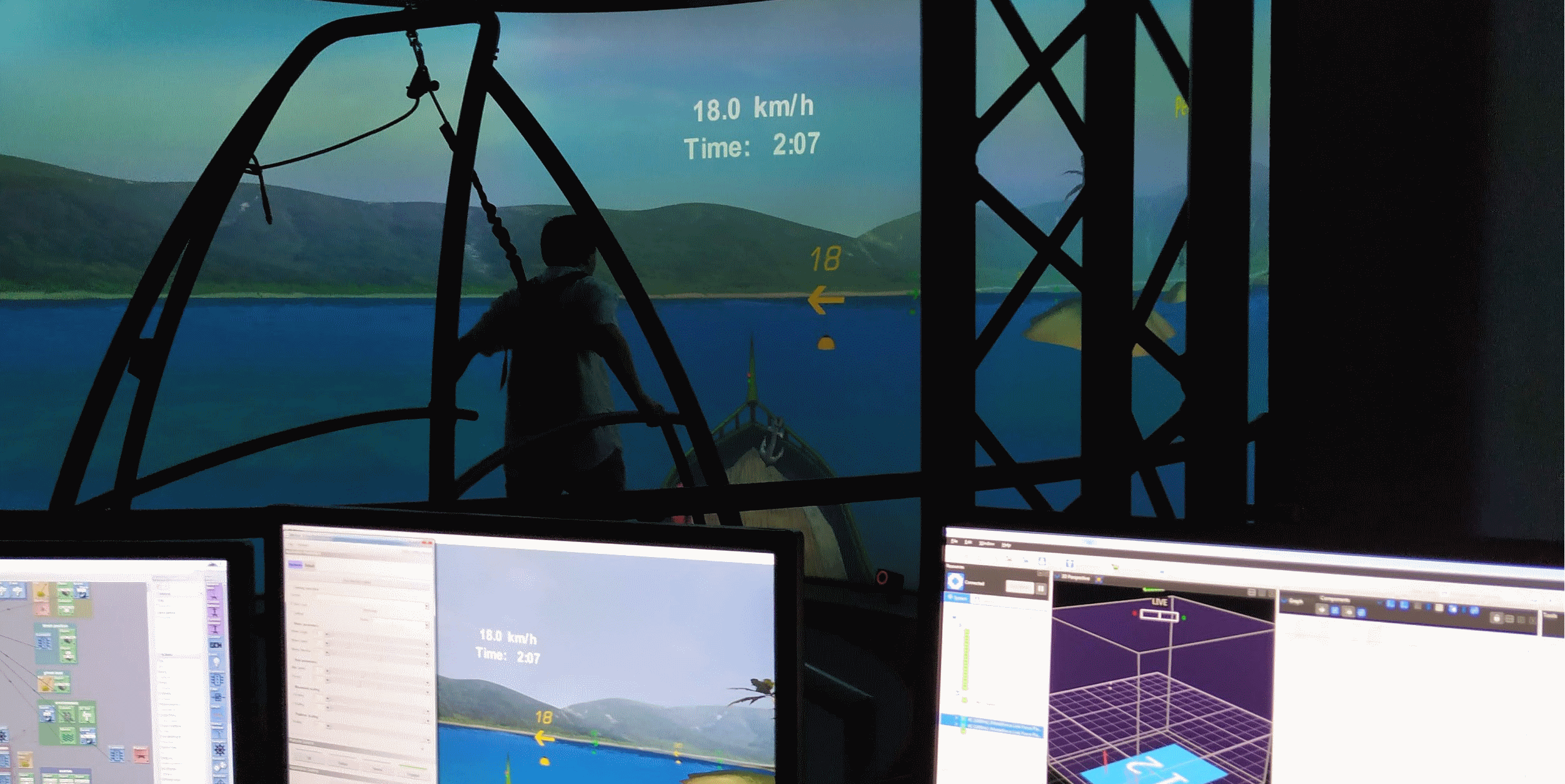}
 \caption{The CAREN system - is a large scale immersive installation with a movement sensitive balance platform and a set of diagnostic tools to track body movement.}
 \label{fig:karen}
\end{figure}

First of all, Virtual Reality is a very broad concept, and it is important to distinguish Immersive Virtual Reality (IVR), usually entered with a head mounted display (HMD), from Virtual Reality (VR) simulated on a traditional computer screen, or in a massive VR installation like CAREN (Figure \ref{fig:karen}).

Grouping insights from these areas together is a somewhat common mistake of novice practitioners caused by the common shorthand of VR for IVR. 

\textbf{Hardware} importance can be also seen in a study \cite{VRcommsfinance2020} which compared videoconferencing via Skype and VR-mediated communication in financial services based on surveys filled out by participants after their two meetings, which were randomly held first either in VR, or via Skype. VR communication achieved significantly higher presence and closeness (to the other person) scores, however, participants rated Skype higher, possibly due to the VR gear being uncomfortable to wear. 

So, even within IVR proper, entered with a HMD, there exist multiple differences which rely on plethora of other factors:

\begin{itemize}

\item Head mounted display: 
while most of the time HMD's specification is publicly available, when it is not, these properties should be reported: per eye or combined resolution, native refresh rate, field of view (horizontal and vertical), display's pixels per inch, display's technology, lenses, eye tracking, recommended user's IPD or IPD range, HMD's weight, tracking system.

\item Controllers: their model information is necessary to assess the ease of use.

\item Graphics processing unit: 
Even simple VR applications can give very poor experience if GPU’s performance is not sufficient to display graphics on two or even three (with on screen reproduction) high resolution screens within target frame rate (e.g. Varjo, Valve Index, Pimax, HTC Vive Pro)\footnote{In our research we observed that dynamic IVR experiences tend to be more immersive with higher refresh rates, which are not critical for non-dynamic HQ IVREs.} 

\item Central processing unit: VR applications tend to be GPU intensive due to combined resolution that needs to be rendered, CPU performance might be a limiting factor when we incorporate various media streaming methods or wireless HMD solutions (e.g. HTC Vive, HTC Vive Pro).

\item Media storage: due to asset loading, it can negatively impact research that relies on e.g. high resolution textures coupled with movement. 

\item Audio accessories: features like active noise cancellation are absent even in high-end VR solutions, so use of external audio solutions should be reported.

\item Cable management systems and extensions: these should be reported as hardware or software when talking about play space area, especially in regards to room scale VR environments. 

If wireless solution was used, information about latency should be provided.

\item External accessories and modifications: Most HMD accessories and modifications alter the factors making up the user experience. 

\end{itemize}

While other PC components are important when configuring systems, their importance from research perspective is little to none as long as complete systems behave as expected. Given that: motherboard chipset, RAM speed, cable extension properties etc, do not need to be reported unless they were in some way the subject of experiment.\\

\textbf{Software} plays an important role in creating coherent experiences and experiments and key affecting solutions should be reported, even if setup reproduction may not be possible:

\begin{itemize}

    \item Framework used: Software like Steam VR, Oculus VR app, proprietary frameworks. 
    These can change HMDs' and controller capabilities.
    
    \item GPU drivers: GPU's are one of the most important pieces in hardware setup for IVR. Different versions can cause a performance boost or instability. 
    
    \item Application engine: Application engine decision is largely based on compatibility, features and team experience. 
    It is important to measure overall performance within our application, to allow for better reproducibility without need for code publication.
    
    \item Boundary system:  Most VR frameworks offer some kind of boundary system, with various levels of customization. Those settings should be noted, especially when exploring immersion-related aspects.
    
    \item User calibration: With ambiguity regarding if users experience content properly, a set of actions should be taken into consideration when fitting equipment to participants. HMD fitting, IPD adjustments, eye tracking calibration are important from experiment validity perspective. Proper vision and comfort validation processes should not only be incorporated into experiments, but also reported if unusual or innovative steps were taken.
    
\end{itemize}

\textbf{Environment} design is another critical factor impacting user experience and immersion.
While environment reproduction might be impossible in derivative works, or when the environment was co-designed in participatory studies \cite{kopec2019vr}, it is important convey enough detail to encourage other researchers to refer to our work, or to become a source of inspiration for them. Sufficient reporting should include:

\begin{itemize}
    
    \item Environment type: Type of environment used - with its features and content - is the most important information. 
    
    \item Interactions: 
    While subset of possible meaningful interactions is defined by environment type and hardware used, it is crucial to describe the subset used in as much detail as possible. Haptic feedback presence and design should be also included in interaction description as it can be highly impactful for certain user groups \cite{h_jakob_2018}. Naturally environment interactivity can be close to none in e.g. 360 videos.
    
    \item Locomotion system: If used one of most important and troublesome aspect of bigger than physical space environments. 
    There is plethora of locomotion techniques to choose from \cite{mti1040024}, not including in-house solutions. 
    Not only we should describe how locomotion was solved but also what complementary systems supported it, such as screen dimming, field of view reductions etc.

    \item Avatar: Usage of certain styles of avatars can increase virtual body ownership and greatly impact immersion.
    Avatar style, body parts' tracking and simulation, and customization options necessitate a detailed description. \cite{waltemate_impact_2018avatar}
    
    \item Control scheme: Control scheme in conjunction with controllers and interactions can make up very unique experiences within identical environments.
    
    \item Graphic fidelity: Graphic fidelity can impact immersion and overall presence, even to the point of increasing emotional response to environments \cite{gamevrealism17}. 
    Its aspects such as graphical style used, content resolution or model complexity should be described, or visually best presented in research description.
    
    \item Sound design: Use of proper sound technology and mixing can lead to more immersive environments and achievement of higher presence. \cite{sound13}
    
    \item Ethical factors: The ethical factors \cite{brey_ethics_1999} conveyed by the environment feel (architectural style, presence of minorities, age groups etc) and interaction design choices (what you can do with items) may be subject to unconscious biases and impact the experience for different groups of users.
    
\end{itemize}

Finally, there is a wide array of possible factors related to \textbf{participants}. 
For example, a comparative study on obedience in VR-mediated communication \cite{VRobedience2019} in which 51 participants were divided into two groups (VR and RR) to solve problems and encouraged to change their answers by the researcher who was with them shows that in the VR setting the participants were more susceptible to manipulation. This effect appears despite the VR avatar of the researcher having limited emotive capacity. However, the study did not account for the participants' familiarity and the perceived presence felt in VR. So there may have been interference caused by lack of experience with VR. Therefore, recruitment strategies and screening for IVR experiments and human factors are critical. Participants can vary considerably, not only in age, socio-economic circumstances, executive and attentional functioning, emotional reactivity but also in how receptive and experienced they are with IVR. 

Some studies show differences in behaviors in Virtual Reality and Real Reality, which can be explained by the extent to which the VR scenarios imitate the RR. Even more differences are found comparing results of one IVR study with another with a different setup. Without a reference to a detailed checklist of factors affecting these experiments conclusions from one study may not be relevant to other studies as the quality of the IVR experience may differ significantly. 

\section{Reference Checklist}

Based on literature review, our experience, practice and a synthesis of key IVR factors we present a reference checklist for describing research conditions of IVR experiments. 

In creating this preliminary checklist we propose a structured way to present IVR papers to include, at least, the following core factors affecting the VR experience and the results:

\begin{enumerate}
\item \textbf{HARDWARE}

\begin{itemize}
\settowidth{\leftmargin}{{\Large$\square$}}\advance\leftmargin\labelsep
\itemsep2pt\relax
\renewcommand{\labelitemi}{$\square$}
\item VR HMD and controllers used
\item Information about GPU, CPU, Media storage
\item Audio sources and inputs (external microphones, headphones with ANC)
\item Wire extensions, cable management systems, wireless solutions (ceiling mounted cables, wireless adapters)
\item Non-standard HMD and controller qualities (lens mods, prescription)
\item Accessories used and modifications (silicone grips, foam replacement)
\end{itemize}

\item \textbf{SOFTWARE}

\begin{itemize}
\settowidth{\leftmargin}{{\Large$\square$}}\advance\leftmargin\labelsep
\itemsep2pt\relax
\renewcommand{\labelitemi}{$\square$}
\item Framework used: Steam VR, Oculus VR or other
\item GPU drivers’ versions
\item Application engine
\item Boundary system and customization
\item User calibration

\end{itemize}

\item \textbf{ENVIRONMENT DESIGN}

\begin{itemize}
\settowidth{\leftmargin}{{\Large$\square$}}\advance\leftmargin\labelsep
\itemsep2pt\relax
\renewcommand{\labelitemi}{$\square$}
\item Environment type
\item Types of interactions
\item Locomotion solutions
\item Avatar look and features
\item Control scheme
\item Graphic fidelity and style illustrated with example assets
\item Sound design
\item Ethical factors and other unique features.

\end{itemize}

\item \textbf{PARTICIPANTS}

\begin{itemize}
\settowidth{\leftmargin}{{\Large$\square$}}\advance\leftmargin\labelsep
\itemsep2pt\relax
\renewcommand{\labelitemi}{$\square$}
\item Recruitment procedures, ethics and informed consent
\item Prior IVR experience and any training received
\item Participant age group and socioeconomic status
\item Exclusions (e.g. participants leaving or finishing early)
\item Glasses/hearing aids
\item VR sickness inducing predispositions (e.g. travel sickness)
\item General attitude towards VR
\item Momentary mood
\item Use of physiological measures of emotional activation (skin conductance)

\end{itemize}

\item \textbf{DISTRACTIONS}

\begin{itemize}
\settowidth{\leftmargin}{{\Large$\square$}}\advance\leftmargin\labelsep
\itemsep2pt\relax
\renewcommand{\labelitemi}{$\square$}
\item Obstacles and problems encountered
\item Waiting times and mean calibration time per participant
\item Other people present during the experiment (e.g. “spotter”)
\item Environmental factors such as wind, smells, temperature and sounds

\end{itemize}

\end{enumerate}

\section{Conclusions}

Relevance of conducted research to other research is of key importance for building a solid foundation of knowledge. For this reason we are excited to present this preliminary checklist that takes into account the specificity of IVR experiments. We hope that this list will be field-tested and expanded according to the diverse experience of the IVR research community. It is our goal to initiate the discussion which would result in a more standardized and comprehensive way of describing IVR experiments, so that researchers with all levels of experience and from very different disciplines can more easily situate the research conditions of each reported IVR experiment on the IVR landscape.

\section{Acknowledgments}
We would like to thank Kobo Association and all transdisciplinary experts involved with the HASE research initiative (Human Aspects in Science and Engineering) including XR Lab at PJAIT, VR Lab at IP PAS, EC Lab at SWPS University and LIT of the NIPI.

\bibliographystyle{splncs04}
\bibliography{bibliography}

\end{document}